\documentclass[twocolumn,english,aps,superscriptaddress,amsmath,amssymb,floatfix]{revtex4-1}
\usepackage[hidelinks]{hyperref}

\usepackage{soul}
\usepackage{gensymb}
\usepackage[utf8]{inputenc}
\usepackage[english]{babel}
\usepackage{amsmath,amsfonts,amssymb}
\usepackage[T1]{fontenc}
\usepackage{url}

\usepackage{cancel}
\usepackage{amsmath}
\usepackage{amsfonts}
\usepackage{amssymb}

\usepackage{graphicx}
\usepackage{epstopdf}

\begin{document} 

\title{Quantifying broadband chromatic drifts in \\Fabry-Pérot resonators for exoplanet science}

\author{Molly Kate Kreider}
\email{mollykate.kreider@colorado.edu}
\affiliation{Department of Physics, University of Colorado Boulder, 440 UCB Boulder, CO 80309, USA}
\affiliation{Electrical Computer and Energy Engineering, University of Colorado Boulder, 425 UCB Boulder, CO 80309, USA}
\affiliation{Time and Frequency Division, National Institute of Standards and Technology, 325 Broadway, Boulder, CO 80305, USA}

\author{Connor Fredrick}
\affiliation{Department of Physics, University of Colorado Boulder, 440 UCB Boulder, CO 80309, USA}
\affiliation{Electrical Computer and Energy Engineering, University of Colorado Boulder, 425 UCB Boulder, CO 80309, USA}
\affiliation{Time and Frequency Division, National Institute of Standards and Technology, 325 Broadway, Boulder, CO 80305, USA}

\author{Scott A. Diddams}
\email{scott.diddams@colorado.edu}
\affiliation{Department of Physics, University of Colorado Boulder, 440 UCB Boulder, CO 80309, USA}
\affiliation{Electrical Computer and Energy Engineering, University of Colorado Boulder, 425 UCB Boulder, CO 80309, USA}
\affiliation{Time and Frequency Division, National Institute of Standards and Technology, 325 Broadway, Boulder, CO 80305, USA}

\author{Ryan C. Terrien}
\affiliation{Department of Physics and Astronomy, Carleton College, One North College Street, Northfield, MN 55057, USA}

\author{Suvrath Mahadevan}
\affiliation{Department of Astronomy \& Astrophysics, 525 Davey Laboratory, The Pennsylvania State University, University Park, PA 16802, USA}
\affiliation{Center for Exoplanets and Habitable Worlds, 525 Davey Laboratory, The Pennsylvania State University, University Park, PA 16802, USA}

\author{Joe P. Ninan}
\affiliation{Dept. of Astronomy and Astrophysics, Tata Institute of Fundamental Research, 1 Homi Bhabha Road, Colaba, Mumbai -400005, India}

\author{Chad F. Bender}
\affiliation{Steward Observatory, University of Arizona, 933 N Cherry Ave, Tucson, AZ 85721, USA}

\author{Daniel Mitchell}
\affiliation{LightMachinery, Inc., 80 Colonnade Rd N, Unit 1, Nepean, ON K2E 7L2, CA}

\author{Jayadev Rajagopal}
\affiliation{NSF NOIRLab, 950 N Cherry Ave, Tucson, AZ 85179, USA}

\author{Arpita Roy}
\affiliation{Schmidt Sciences, New York, NY 10011, USA}

\author{Christian Schwab}
\affiliation{School of Mathematical and Physical Sciences, Macquarie University, Balaclava Road, North Ryde, NSW 2109, Australia}

\author{Jason T. Wright}
\affiliation{Department of Astronomy \& Astrophysics, 525 Davey Laboratory, The Pennsylvania State University, University Park, PA 16802, USA}
\affiliation{Center for Exoplanets and Habitable Worlds, 525 Davey Laboratory, The Pennsylvania State University, University Park, PA 16802, USA}
\affiliation{Penn State Extraterrestrial Intelligence Center, 525 Davey Laboratory, Penn State, University Park, PA, 16802, USA}


\begin{abstract}

The possibility of an Earth-Sun analog beyond our solar system is one of the most longstanding questions in science. At present, answering this question embodies an extremely difficult measurement problem that requires multiple coordinated advances in astronomical telescopes, fiber optics, precision spectrographs, large format detector arrays, and advanced data processing. Taken together, addressing this challenge will require the measurement and calibration of shifts in stellar spectra at the $10^{-10}$ level over multi-year periods. The potential for such precision has recently been advanced by the introduction of laser frequency combs (LFCs) to the field of precision astronomical spectroscopy. However, the expense, complexity and lack of full spectral coverage of LFCs has limited their widespread use and ultimate impact. To address this issue,  we explore simple and robust white-light-illuminated Fabry-Pérot (FP) etalons as spectral calibrators for precise radial velocity measurements. We  track the frequencies of up to 13,000 etalon modes of the installed FPs from two state-of-the-art astronomical spectrographs. Combining these measurements with modeling, we trace unexpected chromatic variations of the FP modes to sub-picometer changes in the dielectric layers of the broad bandwidth FP mirrors. This yields the determination of the frequencies of the FP modes with precision approaching ${\sim} 10^{-11}$/day, equivalent to a radial velocity (RV) Doppler shift of 3 mm/s/day. These results represent critical progress in precision RV measurements on two fronts: first, they make FP etalons a more powerful stand-alone calibration tool, and second, they demonstrate the capability of LFCs to extend cm/s level RV measurement precision over periods approaching a year. Together, these advances highlight a path to achieving spectroscopic calibration at levels that will be critical for finding earths like our own. 
\end{abstract}

\maketitle

\section{Introduction}

\begin{figure*}[!htb]
\centering
\includegraphics[width=\linewidth]{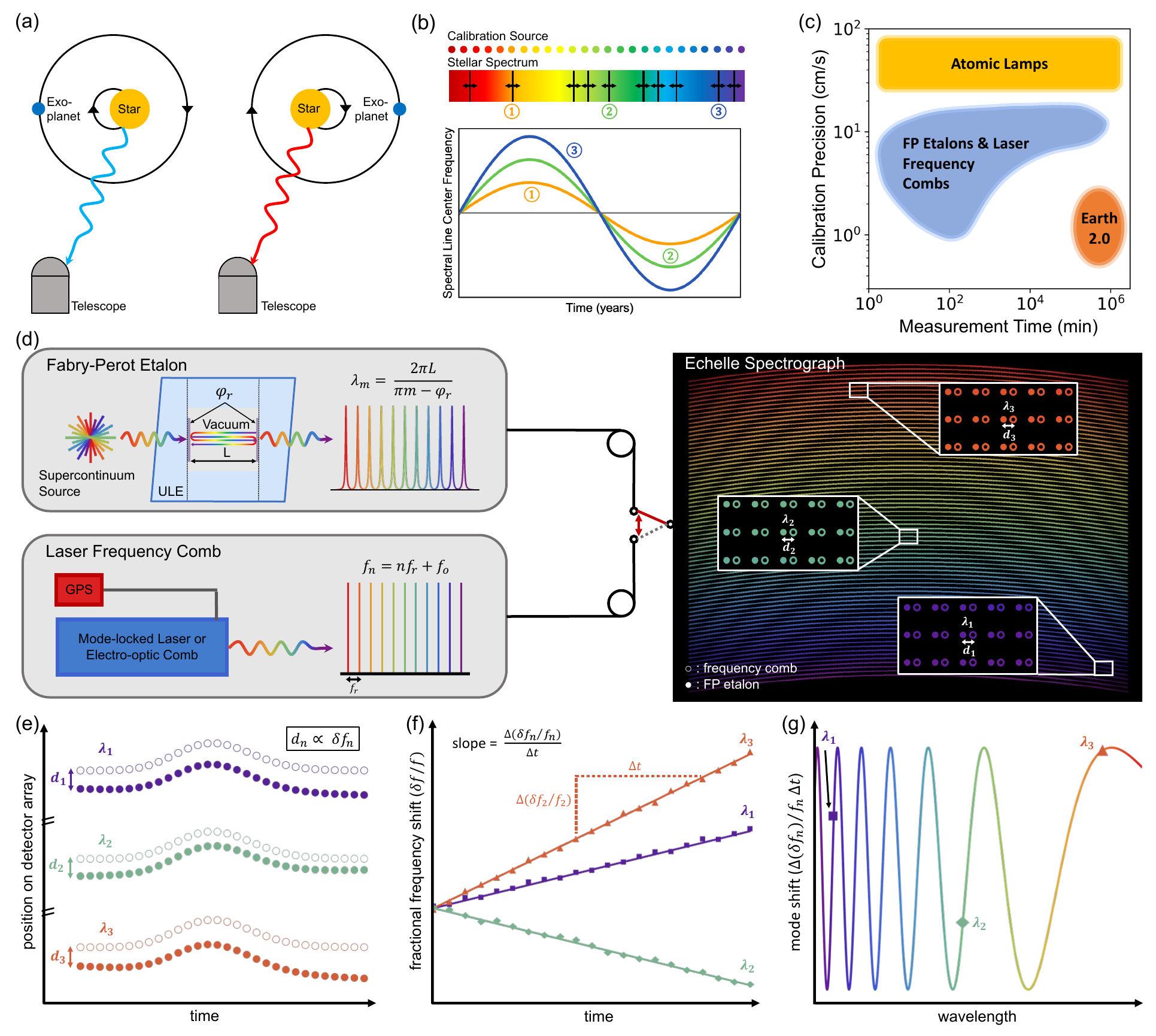}
\caption{(a) Principle of radial velocity exoplanet detection measurement, which uses (b) a known calibration source to measure the periodic Doppler shift of absorption features in a stellar spectrum due to the gravitational pull of an exoplanet. (c) Calibration precision levels needed to detect an Earth-Sun analog, as well as demonstrated precision levels for several different classes of calibrators. (d) Experimental setup of chromatic drift measurements on the HPF and NEID etalons, as detailed in \cite{Terrien2021}. An echelle spectrograph is alternatingly illuminated by laser frequency comb and etalon light. (e) The spatial location (i.e., frequency) of each mode drifts over time, as illustrated in for three modes. (f) The drift of the frequency comb is indicative of instrument drift, and the change in spacing between a frequency comb – etalon mode pair, $\Delta(\delta f_n)$, is indicative of the etalon drift. The fractional frequency shift for any given mode is linear in time. (g) The slopes of the fractional frequency shift can be plotted across the wavelength range of interest, showing the complicated chromatic drift of the etalon \cite{Terrien2021}.}
\label{fig:intro}
\end{figure*}

The discovery and characterization of an Earth-mass planet orbiting a Sun-like star at 1 AU (a.k.a. Earth 2.0) is a long standing challenge in astrophysics--with profound implications regarding the uniqueness of Earth, the formation of planetary systems, and the conditions under which life could exist elsewhere in the Universe. As a result, tremendous effort has been focused on this problem, with the dominant techniques of planetary transits and radial velocity measurements being used in tandem. While a transit, or the passage across the stellar disk, provides information on the size of an exoplanet, measurements of the stellar radial velocity~(RV) provide information on the exoplanet mass through the periodic Doppler shifts of the stellar spectrum (Fig. \ref{fig:intro}a, b). However, RV detection of an Earth analog requires Doppler precision of a few cm/s (fractionally $10^{-10}$) which must be maintained over time scales of multiple years. Ultimately, such detection could be followed by spectroscopic measurements made by instruments like the James Webb Space Telescope to identify biomarkers in the atmosphere.

This is an extremely challenging and multifaceted measurement that requires simultaneous advances along multiple fronts \cite{Fischer_2016}. Presently, some of the most outstanding challenges are due to stellar surface activity and variable telluric contamination from our own atmosphere, both of which look like ``noise'' and mask the desired center-of-mass RV Doppler shift, as well as increase the time and effort it takes to realize long-term measurement cadences with sufficient photon signal-to-noise ratio. Equally importantly are the stable spectrographs and the precise spectrograph calibration that need to provide the stable reference against which tiny Doppler shifts would be measured. Conventional sources for calibration include atomic emission lamps and gas absorption cells, both of which have intrinsic limitations in uniformity and unknown long term stability that hinder their precision \cite{suvrath2012}.

Laser frequency combs (LFCs) provide a broad spectral array of discrete and evenly-spaced emission lines that can be fully stabilized to absolute frequency standards \cite{diddams20}. These properties make LFCs a near ideal spectrograph calibration source \cite{Murphy2007Astrocomb,McCracken:17}, and in recent years, they have shown fractional calibration precision below $10^{-10}$ \cite{Metcalf:19, Probst_2020}. However, they lack full spectral coverage in the blue. Moreover, their implementation remains costly and complex, and is therefore currently limited to a few RV instruments situated at premier facilities. This motivates the use of an alternative calibrator. Moreover, no calibration technique (frequency combs included) has yet demonstrated the capability to resolve 1 cm/s Doppler shifts over timescales of years or longer. As summarized in Fig. \ref{fig:intro}c, there is a gap between the experimentally demonstrated timescales over which etalon- and LFC-calibrated spectrographs have achieved this precision level and the timescales which will be necessary to detect an Earth-Sun analog. 

Our work addresses this issue by exploring new capabilities and resolving systematic uncertainties in Fabry-Pérot etalons, which have long been used in astronomical spectroscopy \cite{Fabry1914}. Broadband, white-light illuminated etalons continue to play an important role in astronomy as a simple, robust calibration source in radial velocity (RV) exoplanet detection. They are used as primary calibrators (in combination with lamps) in some spectrographs (e.g. \cite{Schmidt2021}), and in others, they extend the overall spectral coverage to blue wavelengths not easily achieved with LFCs, but they lack long-term absolute stability.  

We investigate this shortcoming through the analysis and modeling of two extended 6-month cross-calibration studies that resolve puzzling chromatic drifts that have appeared in multiple installed FPs \cite{Terrien2021}. Specifically, we show that, with the help of an LFC, it is possible to track and characterize $\sim~13,000$ modes of a FP with precision that translates to sub-picometer resolution of the optical length of the FP. When combined with careful modeling of relaxation in the FPs dielectric coating, we can disentangle drifts consistent with an achromatic change in cavity length versus higher-order drifts with equivalent RV residuals at a level near 3 mm/s ($10^{-11}$). This new understanding of the complicated drifts of broad bandwidth FP cavities informs future mirror design parameters and immediately improves the achievable precision level when using them as stand alone calibrators that are periodically referenced to absolute frequency standards.  

\section{Measuring chromatic mode drift in FP etalons with LFCs}

Within the landscape of calibration sources, FP etalons are particularly attractive due to their unique combination of high spectral resolution and information content, similar to that of an LFC \cite{Halverson_2014}, but with relative simplicity and cost efficiency. They have become increasingly commonplace in high-precision radial velocity spectrographs, and are used as calibrators in a number of major instruments at large telescopes, including CARMENES \cite{carmenes-fp}, Espresso \cite{schmidt2022chromatic}, HARPS \cite{harps-fp}, HPF \cite{suvrath2012}, KPF \cite{KPF-fp}, MAROON-X \cite{maroonx-fp}, NEID \cite{schwab-2016}, NIRPS \cite{nirp-fp}, and SPIRou \cite{spirou-fp}. They are also slated to be implemented with new instruments to be installed at large telescopes, including iLocator \cite{ilocator} and G-CLEF \cite{gclef-fp}.

While mode-locking fixes the line spacing of an LFC, the modes of an FP cavity are determined by the local resonance condition, which is subject to the wavelength-dependent optical path length of the cavity geometry and mirror coatings. The stability of its spectra is tied to the stability of the cavity's mechanical and optical properties, which can drift with time. Ultra-stable etalons that exhibit a minimal amount of drift have been developed for precision spectroscopy and optical clocks \cite{martin2012high}. Single-frequency near-infrared lasers locked to the modes of silicon resonators held at cryogenic temperatures at a zero crossing of the coefficient of thermal expansion have demonstrated drift rates below one part in $10^{-13}$/day \cite{Storz1998sapphire,Hagemann2014silicon124K,Robinson2019silicon4K}. While not as performant as those designed for cryogenic temperatures, cavities built using room temperature ultra-low expansion (ULE) materials operate in the visible region of the spectrum. Such cavities are still capable of achieving fractional stabilities at $10^{-11}$ from day to day (equivalent to a frequency drift of $\sim0.1$ Hz/s at 1$\mu$m) \cite{Alnis2008ule,Hirata2014ule,Ito2017ule}, and are particularly attractive for astronomical spectroscopy.

One method of circumventing limitations due to drift is to continuously reference one mode of an etalon to an absolute frequency standard, like an atomic transition, and then extrapolate the behavior of that single mode to the behavior of the etalon's entire spectrum \cite{Schwab_2015,Betters:24,Tang_2023}. However, such a technique is only capable of correcting drift in an etalon to the level at which the drift behavior is achromatic. Such a model assumes a fractional change of cavity length $\Delta L/L$ is equal to the fractional frequency change $\Delta f/f$of any cavity mode. This model is applicable when the gradual relaxation of etalon spacer sets the cavity length \cite{Berthold_1977}, and it is well-established that ULE cavity spacers do indeed drift over time \cite{dube_narrow_2009,kobayashi_dissipative_2018,hils_hall_1989}.

However, measurements made at several etalon systems employed in precision radial velocity instruments have shown that etalon mode drift can be significantly and unexpectedly more complex \cite{Jennings:20,Terrien2021,schmidt2022chromatic}. Utilizing LFCs, we have made comprehensive and long-term measurements of the etalons used with the Habitable-zone Planet Finder (HPF) and NEID spectrographs. The unique, broadband nature of these studies, which measure the average daily drift rate of over approximately 5,000 modes (or hundreds of nanometers) of the two FP etalons, is made possible by leveraging the instrument architecture necessary for high precision radial velocity detection.

A simplified schematic of the etalon characterization is illustrated in Figure~\ref{fig:intro}d. An echelle spectrograph (e.g. HPF or NEID) is alternatingly illuminated by the light from a fully stabilized LFC and the temperature-stabilized, vacuum-held ULE etalon \cite{Jennings:20,Terrien2021}. Figure~\ref{fig:intro}e-g isolates three representative pairs of modes from different locations in the spectrum in order to offer an intuitive sketch of the findings of these studies. The wavelengths of the LFC and etalon modes are mapped to relative spatial positions on the spectrograph detector array, where they are monitored over time, as shown in Figure~\ref{fig:intro}e. The overall shared behavior of the wavelengths of etalon and LFC modes is indicative of spectrograph drift. This drift is measured and modeled to high precision using the combined information of all the LFC modes. But it is the change in relative offset between the etalon modes and the wavelength solution from the stable LFC modes that reveals the etalon mode drift. As shown schematically in Figure~\ref{fig:intro}f, in a spectral region ($\lambda_1, \lambda_2, \lambda_3$) the fractional frequency mode drift of the etalon may be linear in time, but the specific slope of the drift varies from one spectral region to another. A qualitative sketch of the behavior of the etalon modes is illustrated in Figure~\ref{fig:intro}g, where the individual data points represent the slope of the drift of etalon modes in a spectral region. 

While illustrated qualitatively in Figure~\ref{fig:intro}g, this oscillatory behavior has been verified for both the NEID and HPF spectrographs, and qualitatively analogous behavior has been found for the etalon system of the ESPRESSO \cite{schmidt2022chromatic} spectrograph. Surprisingly, it implies that the optical length of the etalon is effectively increasing and decreasing in length at different wavelength regions. This poses a clear problem for high-precision RV spectrograph calibration, as such etalon mode drifts are undetectable and thus unaccounted for unless the entire etalon spectrum is continuously compared to a high-precision and broadband absolute frequency reference (i.e., an LFC, but requiring a secondary calibrator defeats the purpose of developing the etalon as a calibrator in the first place). This drift behavior also raises  important considerations for experiments that utilize multiple FP cavity modes for frequency referencing in precision spectroscopy \cite{Hill:21cav,Wang:20cav,Milani:17cav,Arias:18cav,deHond:17cav,bohlouli2006cav,Dawel:24cav,riedle1994cav}, particularly when those modes have large spectral separation.

\section{The role of multilayer coatings in FP drifts}

Previous work has speculated that the wavelength-dependent behavior may stem from the mirror coatings \cite{Jennings:20,Terrien2021}, which are built upon complex optical interferences arising from alternating layers of dielectric materials.  This is one of the least studied aspects of broadband etalon systems. Other work has attempted to benchmark possible drift behavior due to mirror phase drifts \cite{berthold_dimensionality_1976}, but has not been conclusive. Here we have identified the source of the observed chromatic drift behavior in the mirror coatings of the HPF and NEID etalon systems and succeeded in developing a physical model for this drift.

Through our modeling, we determine that the gradual expansion of the outermost layers of the mirror coating causes a mode shift that agrees with the measured HPF and NEID etalon drifts. In particular, we show that the drift behavior of the HPF and NEID etalons can be tied to a $\sim100\times 10^{-15}$ m (100 femtometer) change in the outer layers of the two mirrors. Quantifying these minuscule variations enables us to understand etalon drift behavior at the cm/s level over extended periods. 

\begin{figure*}[!htb]
\centering
\includegraphics[width=\linewidth]{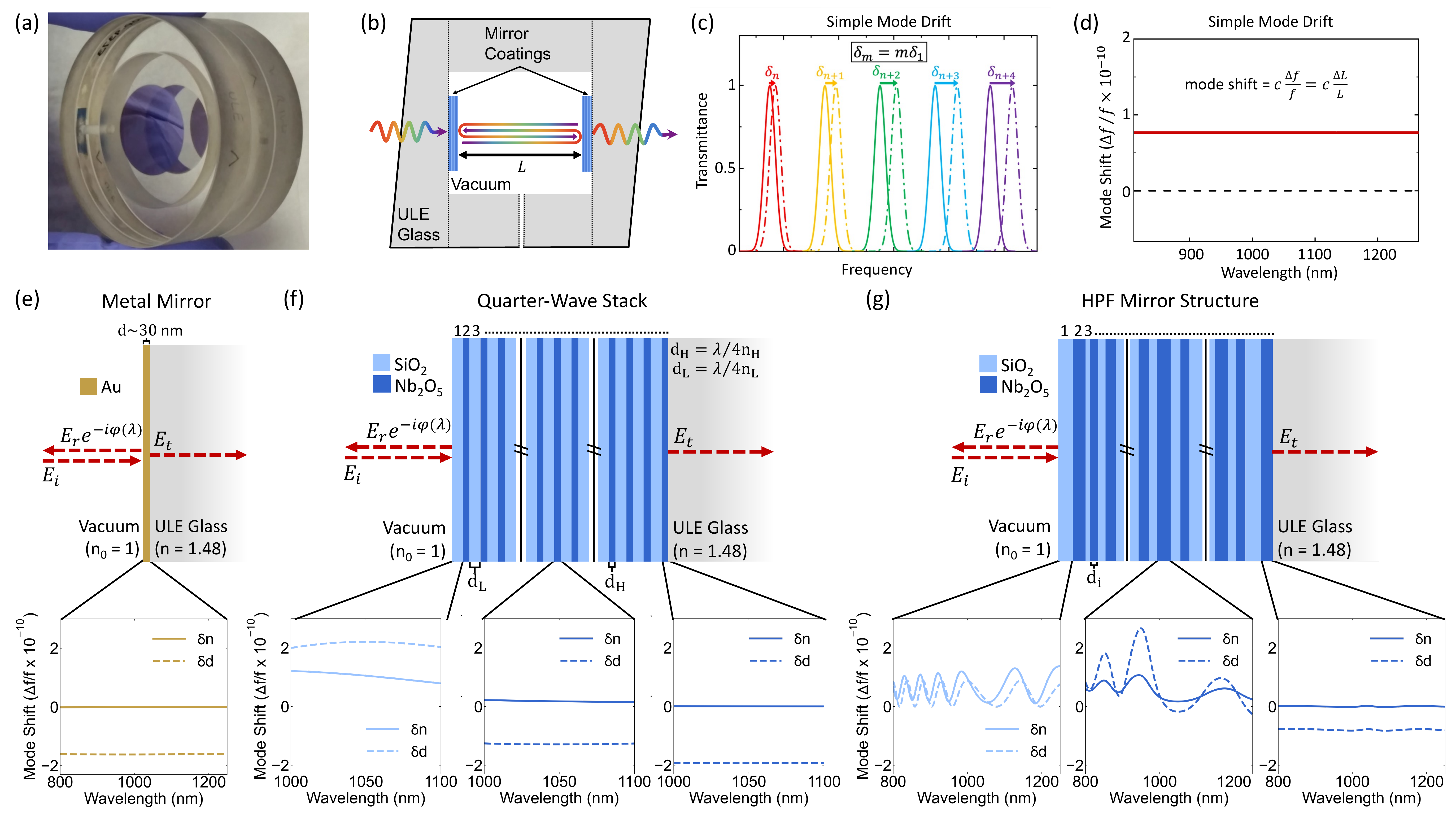}
\caption{Picture (a) and schematic (b) of a Fabry-Pérot (FP) etalon. The simplest drift mechanism in an FP etalon is a change in the spacer length L, which results in the constant, fractional drift behavior shown in (c) and (d). (e-f) illustrate more complicated drift behavior resulting from changes to three mirror structures with 92\% reflectivity. (e) and (f) illustrate the modeled drift resulting from a refractive index ($\delta_n = +10^{-6}$) and layer thickness ($\delta_d = +5 \times 10^{-4}$ nm) change for a thin metal mirror and the top, middle, and bottom layers of a quarter-wave stack, respectively. (g) shows the significantly more complex modeled drift resulting from refractive index ($\delta_n = +10^{-6}$) and layer thickness ($\delta_d = +5 \times 10^{-4}$ nm) changes to the first, sixteenth, and thirty-fourth layer of the HPF mirror structure.}
\label{fig:mir}
\end{figure*}

A Fabry-Pérot etalon, as pictured in Figure~\ref{fig:mir}a and illustrated schematically in Figure~\ref{fig:mir}b, transmits a mode spectrum determined by the resonance condition, which maintains that resonant modes occur at wavelengths where the round trip phase shift of the light in the cavity is equal to an integer multiple of $2\pi$. For an etalon in vacuum, this condition yields modes with frequencies given as
\begin{equation}
f_m = \frac{c}{2 L} (m - \phi_r / \pi).
\label{eq:rc2}
\end{equation}
where $L$ is the cavity length, $c$ is the speed of light, $m$ is an integer, and $\phi_r$ is the reflective mirror phase shift, which is generally wavelength dependent, but assumed to be the same for both mirrors of the etalon. 

It is instructive to consider potential sources of mode drift through this resonance condition, as both $L$ and $\phi_r(\lambda)$ represent physical properties of the cavity that may change over time. The simplest mode drift behavior would stem from a gradual drift in the cavity spacer length, $L$, over time. As has been discussed earlier and as illustrated in \ref{fig:mir}c and \ref{fig:mir}d, this change produces a simple, predictable drift rate across the mode spectrum, with a constant fractional frequency drift that is directly proportional to the fractional spacer change.

In contrast, $\phi_r$ intrinsically varies with wavelength in a complicated manner. We  investigate the impact of various physical perturbations to the coating structure by modeling the shift of $f_m$ using established matrix techniques to determine $\phi_r$ \cite{landrytmm,Kreider2022}. With knowledge of $\phi_r$, we first use a numerical solver to determine a base mode spectrum from Eq.~\ref{eq:rc2}. We then recalculate the mode spectrum using the slightly perturbed $\phi_r$ in order to determine the difference between the new and original mode locations across the wavelength spectrum of interest in units of fractional frequency, $\Delta f/f$. 

Figure \ref{fig:mir}e-g illustrate characteristic modeling results for three different types of mirror coatings with approximate reflectivity of 92\%, which corresponds to that of the mirrors deployed in the HPF and NEID cavities. In each case, individual layers in the coating structure were perturbed by identical changes to their refractive index ($\delta_n = +10^{-6}$) and layer thickness ($\delta_d = +5\times10^{-4}$ nm).

Figure \ref{fig:mir}e illustrates mode drift for the simplest possible mirror, a thin layer of metal. As one may expect due to the relative simplicity of a single layer, perturbing a 30 nm gold layer produces minimal drift across the wavelength spectrum. Specifically, perturbing the refractive index has effectively no impact on the mode spectrum, and, as our model assumes that the gold layer expands into the cavity, the impact from the layer thickness perturbation stems primarily from the effective spacer length decrease. Unfortunately, these mirrors are in general not suitable for cavities deployed for astronomical applications due to the metal's absorption and resulting low transmission on resonance.

Figure \ref{fig:mir}f illustrates mode drift for a simple quarter-wave stack comprising alternating layers of silica and niobia. The reflected phase from a quarter-wave stack is fairly simple, and this simplicity means that perturbations to the coating structure do not give rise to the chromatic drift observed in the HPF and NEID etalons. However, such mirrors are generally unsuitable for broadband spectrographs like HPF and NEID, as they do not have a flat reflectivity spectrum across a sufficiently broad bandwidth. 

The mirror coatings used in the HPF and NEID etalons are deposited via magnetron sputtering and comprise alternating layers of silica and niobia (with the outermost layer, i.e., the layer closest to vacuum, which we refer to as the "first" layer, being silica in both cases). They are significantly more complicated, as having a constant reflectivity curve across hundreds of nanometers of bandwidth requires large variations in layer thickness. A simple schematic of the model and results is shown in Figure \ref{fig:mir}g for the 34-layer HPF structure; the 48-layer NEID etalon exhibits qualitatively similar behavior. We find that perturbations to individual layers of the mirror structure can be broken into regions with qualitatively similar behavior. From this analysis, we  identify three general, qualitative categories of interest: the outermost < 5 layers, the middle layers, and the deepest ~10 layers. All layers include an offset caused by the effective change in cavity length due to the change in layer thickness. Changes to the outermost five layers of the mirror are similar in their frequency of oscillation and have a relatively consistent amplitude across the wavelength spectrum. These qualitative features exhibit close agreement with the measured behavior of the HPF and NEID etalons. Changes to the middle layers of the mirror exhibit comparably and increasingly (with layer number) fewer oscillations over the wavelength range of interest. The characteristic drifts associated with the middle layers have a reasonably large magnitude but are inconsistent with the measured drift behavior. Due to the non-negligible magnitude, were these layers changing, their effect would be readily observable in the measured behavior. Therefore, we can conclude that the middle layers are not drifting due to being qualitatively inconsistent with the measured behavior. Changes made to the deepest layers of the system similarly exhibit little oscillatory behavior and have much smaller magnitudes of effect relative to the other layers.

\section{Observations and Data}
HPF and NEID are both stabilized, fiber-fed, cross-dispersed echelle spectrographs that are optimized for RV exoplanet detection. HPF, located at the Hobby-Eberly Telescope at McDonald Observatory in Texas, operates in the near-infrared (820-1280nm) at a spectral resolution ($\lambda / \Delta\lambda$) of approximately 55,000 \cite{suvrath-2014}. NEID, located at the WIYN telescope at the Kitt Peak Observatory in Arizona, operates at visible wavelengths (380-930nm) at a spectral resolution of 110,000 \cite{halverson-2016, schwab-2016, robertson-2019}. Both systems are equipped with a suite of calibration sources including laser frequency combs (LFCs) that provide a reliable absolute calibration, as well as Fabry-Perot etalons (FPs) which are the devices of interest here. Spectra of these sources are recorded daily, enabling the long-term tracking of the relative, wavelength-resolved drift of their spectral features. 

The FP system installed at HPF was described in detail in Terrien et al. 2021. The FP system at NEID is similar to the one at HPF.  Planar dielectric mirrors manufactured by Light Machinery are attached to a 7.5mm ULE spacer, and sit in a stabilized chamber designed by Stable Laser Systems. Light from an NKT SuperK Extreme supercontinuum source is coupled to the FP via single-mode fiber (LMA-5 fiber from NKT). Finesse (F>~30) of the FP is maintained across the NEID bandpass, motivating the many-layer mirror structure described above. This system delivers to NEID a mode spectrum of distinct, 20-GHz-spaced modes that can be used for spectrograph calibration.

For this study, we examined multi-month FP/LFC relative drift monitoring from both HPF and NEID. The HPF spectral dataset, along with associated methods for measuring the mode centroids and data quality, are described in Terrien et al. 2021. The same analysis pipeline was used to analyze spectra from the NEID FP system, and we briefly summarize its application to the NEID spectra here. We focused on ~1350 FP spectra spanning 2021-5-1 to 2021-8-1, which we downloaded from the NEID archive (\url{https://neid.ipac.caltech.edu/search.php}). These spectra were obtained through the CAL fiber in HR mode. Spectra in the NEID archive are provided in wavelength-calibrated form by the standard NEID pipeline (v1.2.1), which uses the LFC in combination with the FP and hollow cathode lamps to map pixel index to wavelength \footnote{The NEID pipeline documentation describes the details of this process \url{https://neid.ipac.caltech.edu/docs/NEID-DRP/}}. For this work, we focused on the wavelength range most reliably spanned by the LFC (5100-9300A) where the wavelength calibration is of the highest quality. This waveband contains more than 13,000 FP modes. As for HPF \cite{Terrien2021}, we measured the wavelength centroid of each of nearly 5,000 FP mode using a least-squares fit of a Gaussian function, and tracked the motion of these centroids over time. Over the time span studied, the behavior of the FP modes are generally linear, so we quantified the wavelength-dependent drift using a least squares fit of a line to the time series of centroids for each mode.

\section{Modeling chromatic mode drift in HPF and NEID FPs}

\begin{figure*}[!htb]
\centering
\includegraphics[width=14cm]{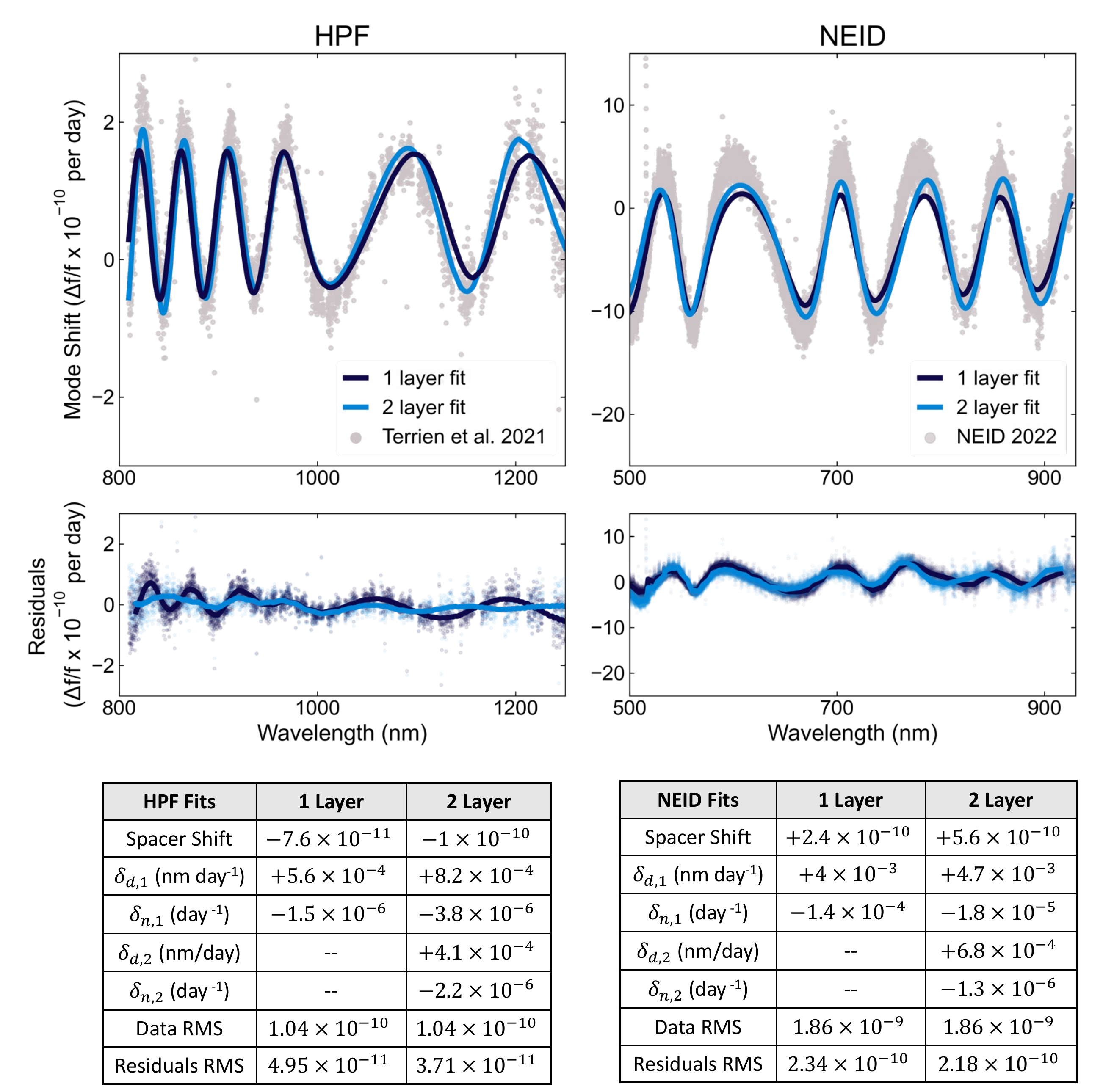}
\caption{Fits of the perturbative models to the measured HPF etalon data \cite{Terrien2021} and 2022 NEID etalon data. Residuals of the model fits are shown in the lower panels for both etalons, with lines corresponding to the moving averages. Single and double layer fit parameters and values for HPF and NEID are recorded in the tables below, along with the RMS values of the data pre-fit and RMS values for the model residuals.}
\label{fig:fits}
\end{figure*}

A full analysis of the two etalons is presented in Figure~\ref{fig:fits}, which shows fits for changes to the outermost layer and the outermost two layers for the HPF and NEID etalons. These results are calculated as a linear combination of the effect of changes to individual layer thicknesses or refractive indices (i.e., the functions shown in Figure \ref{fig:mir}). There are two fit parameters for each layer in a given model: the first for the refractive index and the second for the thickness change of each layer. In addition, we include a fit parameter to account for the bulk offset due to gradual spacer relaxation. Although fit independently and without constraints, the relative changes to the refractive index and layer thickness are consistent (within a factor of two) with simple models that couple refractive index to changes in density \cite{burnett_1927}. Fits including more than the outermost two layers do not yield reasonable parameters when constrained to physically relevant orders of magnitude and relative sign. For example, a three layer fit models a refractive index change in the third layer with no change to the same layer's thickness.

Structure in the residuals of the single-layer HPF and NEID fits may be due to differences between the layer thicknesses input into the model and those that were actually manufactured. The single layer HPF fit of Figure~\ref{fig:fits}a, is slightly wavelength shifted from the measured data. Manufacturing tolerances are expected on the deposited layer thicknesses, and we have found that randomly perturbing the entire structure shifts the relative wavelength position of the single layer models. For this reason, while it is reasonable to conclude that changes in the first few mirror layers are the likely cause of the measured behavior of the HPF and NEID etalons, fully determining the mechanism and drift rates would require further study of the exact physical properties of the mirror. Specifically, we anticipate that it will be important to have more accurate information on the refractive index and deposited thickness of each layer than we currently possess.

\begin{figure}[!htb]
\centering
\includegraphics[width=\linewidth]{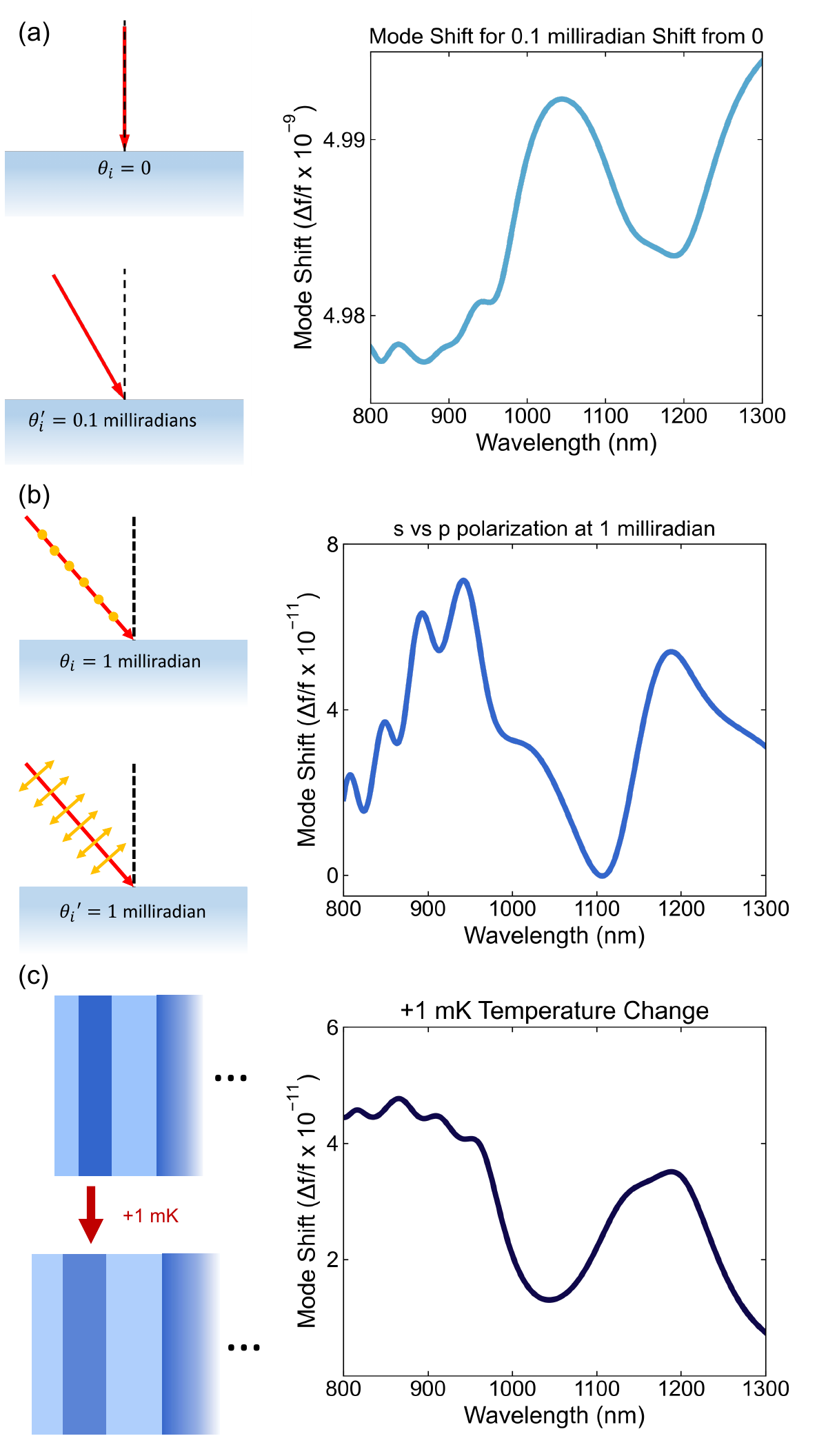}
\caption{Other perturbations to HPF etalon system, all of which are unlikely sources of measured drift. (a) Modeled fractional frequency shifts due to a 0.1 milliradian angular shift from normal incidence. The dominant effect of the mode shift is a large vertical offset due to a change in the effective cavity spacer length, with wavelength variations smaller by a factor of 103. (b) Modeled fractional frequency shifts between the modes of s and p polarized light at 1 milliradian incidence. The characteristic shape bears some qualitative resemblance to the HPF results; however, achieving the scale of the measured HPF behavior would require an unlikely combination of large angular misalignment and long-term monotonic polarization drift. (c) Modeled fractional frequency shifts due to a uniform 1 mK change across the mirror stack. Changes involving the entire mirror stack do not show agreement with the measured data due to the contribution of the middle layers.}
\label{fig:othertests}
\end{figure}

While the primary drift mechanism of the HPF and NEID etalons can be explained by perturbations to the outermost layers of the mirror stack, we also conducted a thorough investigation into other physically reasonable drift mechanisms, including polarization and incident angle changes, temperature changes, and manufacturing tolerances. These modeling results are summarized for the HPF etalon in Figure~\ref{fig:othertests}. While not shown here, analysis of the layer by layer drift functions for the NEID etalon yields that only the outer layers bear any qualitative resemblance to the measured drifts. We have therefore concluded that none of these alternative mechanisms are responsible for the etalon drift measured by the HPF or NEID spectrographs. However, these null tests, which could still be present in the residuals of the HPF and NEID etalons, or in other systems more generally, provide a useful catalogue for potential characteristic drift behaviors in etalons.

The HPF etalon is aligned so that light in the cavity propagates normal to the mirror surfaces. However, it is not unreasonable to assume that there may be some variation in the alignment over time, particularly were that misalignment in the range of 0.1 milliradians from normal \cite{Jennings:20}. The effect of such a change in alignment is shown in Figure~\ref{fig:othertests}a. In this figure, it is clear to see that the dominant effect is an offset in the mode shift due to the effective cavity length change from the longer propagation distance ($\frac{L}{\cos\theta}$ rather than simply $L$). Variations across the wavelength spectrum are several orders of magnitude smaller than this offset, meaning that the chromatic behavior in the HPF etalon is not driven by drift of the incident angle.

At normal incidence, the Fresnel coefficients for a single boundary converge, and there is no difference between the phase shift of $s$ and $p$ polarized light. However, even at very small incident angles, the resonant wavelengths for the two polarization components begin to diverge. As shown in figure~\ref{fig:othertests}b, at 1 milliradian incidence the difference between the $s$ and $p$ modes is less than 1 part in $10^{-10}$. In order for a small change in polarization to affect mode drift on the order of magnitude seen in the HPF etalon, the required angular misalignment would be $\sim$ 10 milliradians. This is well beyond the alignment and assembly tolerances, making polarization fluctuation an unlikely factor in the HPF results.

The HPF etalon is held in vacuum at $~32.5\degree$C with a temperature precision approaching $1$~mK. SiO$_2$ and Nb$_2$O$_5$ are not particularly temperature sensitive: SiO$_2$ has a coefficient of thermal expansion (CTE) of $~0.55\times10^{-6}$~K$^{-1}$ and thermal refractive index of $~11\times10^{-6}$~$\degree$C$^{-1}$ \cite{Malitson:65,Amotchkina:17}, and Nb$_2$O$_5$ has a CTE of $~5.8\times10^{-6}$~K$^{-1}$ and thermal refractive index of $~14\times10^{-6}$~$\degree$C$^{-1}$ \cite{chen}. The results of a 1 mK temperature change across the entire mirror structure is shown in Figure~\ref{fig:othertests}c and yields a mode shift on the order of magnitude of the measured mode behavior, although the shape does not show many qualitative similarities with the chromatic variation of the measured drift.

Aside from the issues raised above, and most importantly, we note that none of these additional perturbations exhibit close qualitative agreement with the HPF measurements, making them unlikely sources of mode drift. We also emphasize that the etalon measurements shown in Figure~\ref{fig:fits} are of the average drift per day, and it is unlikely that random fluctuations of the angular misalignment, polarization, or temperature would contribute to the mode shifts over time in such a regular, consistent fashion.

\section{Discussion and future work}
In summary, we have modeled the effects of a number of mirror-coating related causes for the chromatic mode drift in the HPF and NEID etalons, including incident angle variation, polarization fluctuation, and temperature gradients. We are able to exclude the majority of these simple changes as the mechanism responsible for the HPF and NEID mode drifts. We also have approached the modeling via a perturbative layer-by-layer analysis of the mirror coating stack. Via this approach, we found that changes to the first few layers of the mirror system cause mode shifts that agree with the observed HPF and NEID etalon behavior. The most likely mechanism for the wavelength-dependent mode shift is a gradual expansion of the first one or two mirror coating layers over time. A reasonable explanation for this behavior comes from the fact that the magnetron-sputtered stacks have some level of intrinsic, compressive stress due to the deposition method \cite{liu_elastic_1992,abadias_situ_2008,thornton_internal_1979,wu_intrinsic_1979,hong_aging_1985}. The outermost layer of the structure faces vacuum, which offers little resistance, meaning that this layer can gradually expand and relax (where as other, deeper layers in the system are more constrained by the surrounding layers and substrate, which offer some resistance to such a process). This relaxation could produce refractive index and thickness changes that are consistent with the relative magnitude and sign of the parameters fit by the models discussed above. It is also possible that the drift behavior could be explained by the deposition of some contaminant in the vacuum chamber (e.g., water) on the mirror surfaces. The deposition of some other material on the mirror stack is effectively equivalent to a change in the outermost layer of the stack, so our model is unable to distinguish between the two. We further note that neither of these explanations covers FPs with soft coatings (e.g., ESPRESSO), as soft coatings are not compressive and are more prone to processes like outgassing or water absorption. 

The modeling results we present are by nature specific to the HPF and NEID etalons, and further analysis of other etalon systems will be necessary to determine the dominant effects of the coatings in each. However, they do give clear evidence for etalon mirror coatings as the mechanism behind wavelength-dependent mode drift. As such, our work lays the foundation for additional modeling, mirror design and fabrication with the goal of identifying more stable mirror coatings for future etalon systems. In the design of such etalon mirrors, it is most typical that only the bandwidth and reflectivity are optimized. With the addition of our insights and modeling, it should now be possible to impose additional constraints in the coating design. In particular, we note that the complexity of the mirror phase across the wavelength region of interest is directly tied to the etalon drift behavior for HPF and NEID (specifically, in both cases, the drift behavior is proportional to $\cos \phi_r$). Therefore, constraining mirror designs to phases with smaller variations in $\cos \phi_r$ across the spectrum should mean that, even for the drift mechanisms explored in this paper, the drift behavior would be less complex. This would in turn make locking the cavity to a known reference more straightforward. In general, in this largely unexplored high-dimensional design space, we expect there to emerge opportunities that minimize the impact of coating structure, aging, and manufacturing tolerances to provide more stable and predictable etalon behavior.

Finally, we note that the broadband and high experimental precision required to understand the drift mechanisms of Fabry-Pérot etalons at the cm/s level and at > 6 month timescales is not dissimilar to the demands of extreme precision radial velocity measurements. Our work thus also separately provides a unique and critical validation of laser frequency combs as astronomical calibrators, offering clear evidence that they are capable of performing at both the precision level and timescale that will be required for the detection of an Earth analog. In seeking to understand, characterize, and improve our one of our oldest calibration sources, we therefore simultaneously offer a crucial validation of our newest ones. 

\section{Acknowledgements}
The authors acknowledge helpful discussions with Julian St\"urmer on mirror coatings, who has also pointed out the importance of relaxation of the stress in the outer dielectric coating layers and its impact on the variation of the mirror properties

The authors are also grateful for financial support from NIST, award 70NANB18H006 from the U.S. Department of Commerce, and from NSF grants AAG 2108512, ATI 2009889, ATI 2009982, ATI 2009955, ATI 1310875, and ATI 1310885, ATI 2009889, ATI 2009982
The Hobby-Eberly Telescope (HET) is a joint project of the University of Texas at Austin, the Pennsylvania State University, Ludwig-Maximilians-Universität München, and Georg-August-Universität Göttingen. The HET is named in honor of its principal benefactors, William P. Hobby and Robert E. Eberly. The Center for Exoplanets and Habitable Worlds and the Penn State Extraterrestrial Intelligence Center are supported by Penn State and its Eberly College of Science.

\bibliography{sample}

\appendix
\section{Transfer Matrix Method}

We begin by assuming that the electric fields in the incident and transmitted media can be related by a matrix M, such that
\begin{equation}
\begin{pmatrix} E_i\\E_r \end{pmatrix} =
\begin{pmatrix} M_{11} & M_{12}\\ M_{21} & M_{22}\end{pmatrix} \begin{pmatrix} E_t\\0 \end{pmatrix}.
\label{eq:me}
\end{equation}
M can be calculated as the net product of a series of boundary ($B$) and propagation ($P$) matrices for each defined boundary $j, j+1$ and finite layer $j$ of the system, where the boundary matrices are related to the reflection and transmission at that interface and the propagation matrices describe the phase shift picked up by the light as it propagates across the corresponding layer. For an $n$ layer system, then, $M$ is given by
\begin{equation}
M=B_{1,2}P_2 B_{2,3}...P_{n-1}B_{n-1,n}.
\end{equation}
Mathematically, these matrices can be written as
\begin{equation}
B_{j,j+1}=\frac{1}{t^{s|p}_{j,j+1}}
\begin{pmatrix} 1 & r^{s|p}_{j,j+1}\\ r^{s|p}_{j,j+1} & 1\end{pmatrix},
\end{equation}
where $r^{s|p}_{j,j+1}$ and $t^{s|p}_{j,j+1}$ denote the Fresnel reflection and transmission coefficients at the boundary between layer $j$ and $j+1$, respectively, for either $s$ or $p$ polarized light, and
\begin{equation}
P_{j}=\begin{pmatrix} e^{-i \phi_j}& 0\\ 0 & e^{i \phi_j}\end{pmatrix},
\end{equation}
where $\phi_j$ is the phase shift in a single layer of refractive index $n_j$ and thickness $d_j$,
\begin{equation}
\phi_{j}=\frac{2\pi}{\lambda} n_j d_j \cos{\theta^{t}_{j}},
\end{equation}
with $\lambda$ representing the incident wavelength and $\theta_{j}^{t}$ representing the angle at which the light is transmitted into the layer.

From Eq.~\ref{eq:me}, we can define the complex reflection coefficient for the system, $r$, as
\begin{equation}
r=\frac{E_r}{E_i}=\frac{M_{21}}{M_{11}}.
\end{equation}
The mirror phase shift can then be extracted from this quantity via the relationship
\begin{equation}
\phi_r=\arctan{\frac{\operatorname{Im}(r)}{\operatorname{Re}(r)}}.
\end{equation}

\end{document}